\documentclass[10pt,twocolumn,floatfix,prl,superscriptaddress]{revtex4-1}
\usepackage{amsmath,amssymb,amsthm,mathrsfs,amsfonts,dsfont,amstext} 
\usepackage{textcomp,pbox}
\usepackage[export]{adjustbox}
\usepackage{bm}
\usepackage{dcolumn,booktabs,url}
\usepackage[scaled]{helvet}
\usepackage{sansmath,gensymb}
\usepackage{tikz,graphicx,transparent,color}
\usepackage{multirow}
\usepackage[separate-uncertainty = true]{siunitx}
\usepackage{comment}
\usepackage{physics}
\usepackage{pdfpages}
\usepackage{array}

\usepackage{pdfpages}

\makeatletter
\AtBeginDocument{\let\LS@rot\@undefined}
\makeatother

\newcolumntype{C}[1]{>{\centering\let\newline\\\arraybackslash\hspace{0pt}}m{#1}}

\usepackage[colorlinks=true]{hyperref}
\usepackage{graphicx}

\hypersetup{
     colorlinks   = true,
     citecolor    = red,
     linkcolor    = blue,
     urlcolor     = red     
}

\graphicspath{{./figs/}}

\newcommand{\SM}{Supplemental materials}
\newcommand{\yso}{Y$_2$SiO$_5$}

\newcommand{\ybi}[0]{$^{171}$Yb$^{3+}$}

\newcommand{\ybiso}[0]{$^{171}$Yb$^{3+}$:Y$_2$SiO$_5$}

\newcommand{\euiso}[0]{$^{151}$Eu$^{3+}$}

\newcommand{\transition}{$^2$F$_{7/2}(0) \longleftrightarrow ^2$F$_{5/2}(0)$ }
\newcommand{\gstate}{$^2$F$_{7/2}$}
\newcommand{\estate}{$^2$F$_{5/2}$}
\newcommand{\figref}[1]{\figurename{~\ref{#1}}}

\begin{document}

\newcommand{\TitleName}{Optical spin-wave storage in a solid-state hybridized electron-nuclear spin ensemble}
\title{\TitleName}

\newcommand{\AffGeneve}{Department of Applied Physics, University of Geneva, CH-1211 Gen\`{e}ve, Switzerland}
\newcommand{\AffParis}{Chimie ParisTech, PSL University, CNRS, Institut de Recherche de Chimie Paris, 75005 Paris, France}
\newcommand{\AffPariss}{Facult\'e des Sciences et Ingénierie,  Sorbonne Universit\'e, UFR 933, 75005 Paris, France}

\author{M.~Businger}
\thanks{These authors contributed equally to this work.}
\affiliation{\AffGeneve{}}
\author{A.~Tiranov}
\thanks{These authors contributed equally to this work.}
\altaffiliation{Present address: The Niels Bohr Institute, University of Copenagen, DK-2100 Copenhagen $\oslash$, Denmark}
\affiliation{\AffGeneve{}}
\author{K.~T.~Kaczmarek}
\affiliation{\AffGeneve{}}
\author{S.~Welinski}
\altaffiliation{Present address: Department of Electrical Engineering, Princeton University, Princeton, NJ 08544, USA}
\affiliation{\AffParis{}}
\author{Z.~Zhang}
\affiliation{\AffParis{}}
\author{A.~Ferrier}
\affiliation{\AffParis{}}
\affiliation{\AffPariss{}}
\author{P.~Goldner}
\affiliation{\AffParis{}}
\author{M.~Afzelius}\email[Email to: ]{mikael.afzelius@unige.ch}
\affiliation{\AffGeneve{}}

\date{\today}

\begin{abstract}

Solid-state impurity spins with optical control are currently investigated for quantum networks and repeaters. Among these, rare-earth-ion doped crystals are promising as quantum memories for light, with potentially long storage time, high multimode capacity, and high bandwidth. However, with spins there is often a tradeoff between bandwidth, which favors electronic spin, and memory time, which favors nuclear spins. Here, we present optical storage experiments using highly hybridized electron-nuclear hyperfine states in \ybiso{}, where the hybridization can potentially offer both long storage time and high bandwidth. We reach a storage time of 1.2 ms and an optical storage bandwidth of 10 MHz that is currently only limited by the Rabi frequency of the optical control pulses. The memory efficiency in this proof-of-principle demonstration was about 3\%. The experiment constitutes the first optical storage using spin states in any rare-earth ion with electronic spin. These results pave the way for rare-earth based quantum memories with high bandwidth, long storage time and high multimode capacity, a key resource for quantum repeaters.
\end{abstract}

\maketitle 

Solid-state impurity spins play an increasingly important role in quantum information technologies, with applications in communication, computing and sensing~\cite{Bussieres2014,Degen2017,Awschalom2018}. In quantum communication, solid-state spins that can be interfaced with photons are promising candidates for nodes in quantum networks and quantum repeaters. In such systems, optical transitions are used to convert quantum information between internal spin states and optical photons, where spins store and possibly process quantum information within the solid~\cite{Atatuere2018,Awschalom2018}. 

There is a current interest in using both the electronic and nuclear degrees of freedom of solid-state spin systems~\cite{Awschalom2018,Bradley2019}, with the goal of simultaneously achieving efficient manipulation and long-duration storage. Electronic spins couple strongly to external fields, making them ideal for high-bandwidth operations and highly sensitive sensors. The weaker coupling of the nuclear spins shields them from the environment, allowing long-duration quantum storage. However, finding spin systems that simultaneously possess good optical, electronic and nuclear spin properties is challenging. A prominent example is the nitrogen vacancy (NV) center in diamond, where the electron spin of the NV can be coupled to photons~\cite{Doherty2013}, while the hyperfine interaction with neighbouring $^{13}$C nuclear spin provides a long-duration memory~\cite{Maurer2012,Bradley2019}. Similar hybrid electron-nuclear systems are investigated using $^{31}$P phosphor donors in silicon~\cite{Steger2012,Saeedi2013} and quantum dots~\cite{Gangloff2019}. 

\begin{figure*}
	\includegraphics[width=\linewidth]{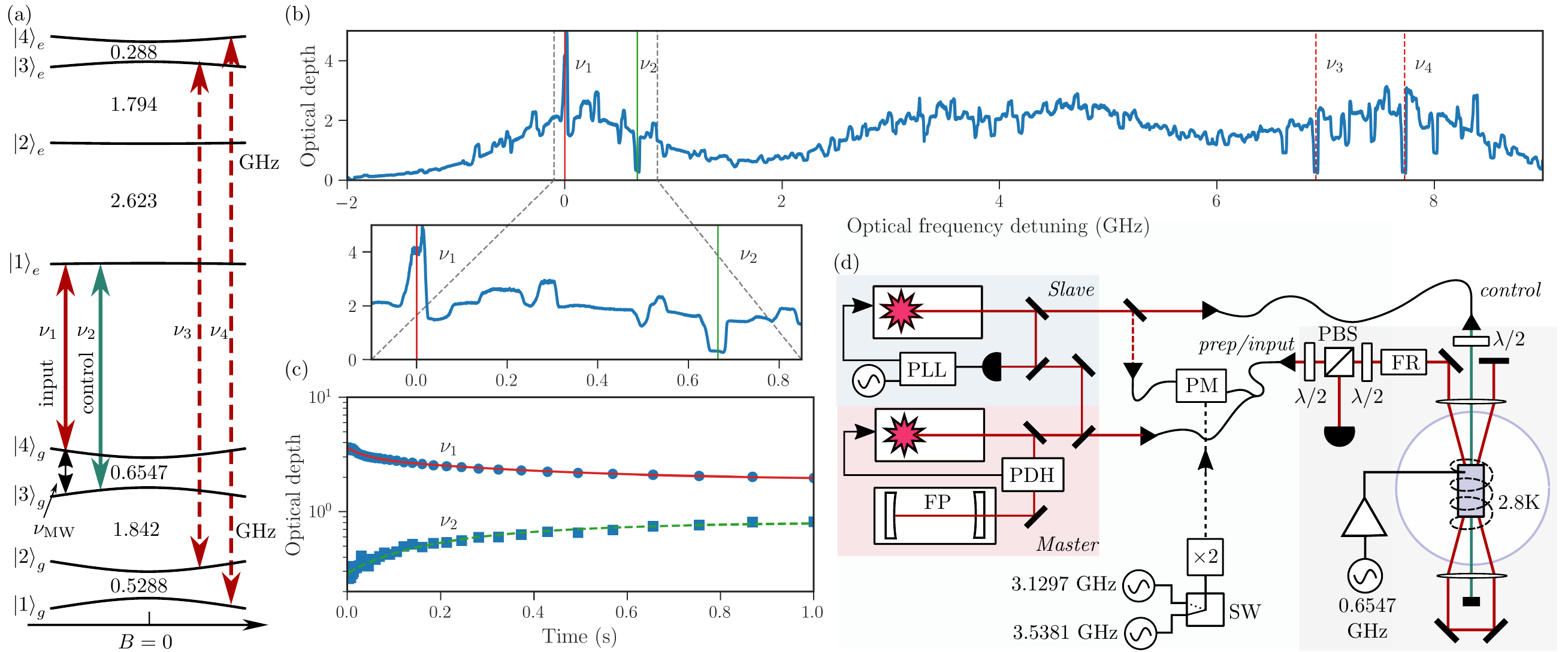}
	\caption{(color online) \textbf{Experimental setup}. 
		\textbf{(a)} Energy level diagram of the optical \transition{} transition for site II of \ybiso{} crystal at zero magnetic field. The $\Lambda$ system used for optical storage uses transition $\nu_1$ for the input/output pulses and transition $\nu_2$ for the control pulses, and $\nu_{\text{MW}}$ for the MW pulses. Transitions $\nu_3$ and  $\nu_4$ are added for the different optical pumping steps.
		\textbf{(b)} The absorption spectrum after performing class cleaning and state initialization to the $\ket{4}_g$ ground state over a 40~MHz bandwidth. Inset: zoom on the $\nu_1$ and  $\nu_2$ transitions of the $\Lambda$ system. 
		\textbf{(c)} Optical depth of the anti-hole at $\nu_1$ frequency and the hole at $\nu_2$ frequency, as a function of delay after the state initialization. 
		\textbf{(d)} Experimental setup (see text for details). FP - Fabry-Perot cavity, PDH - Pound-Drever-Hall module, PM - phase modulator, PLL - phase locked loop, $\lambda/2$ - half-wave plate, PBS - polarization beamsplitter, , FR - Faraday rotator, $\times 2$ - microwave frequency doubler, SW - microwave switch. 
		}
	\label{fig:exp_setup}
\end{figure*}

Rare-earth-ion (RE) doped crystals represent another prominent example of solid-state impurities with excellent optical~\cite{Thiel2011} and spin properties~\cite{Zhong2015}. RE crystals have emerged as strong candidates for ensemble-based quantum memories (QMs) \cite{Hedges2010,Clausen2011,Saglamyurek2011,Zhou2012}, and more recently for single ion/spin~\cite{Kolesov2012,Dibos2018,Zhong2018,Casabone2018} quantum information processing. For ensemble-based QMs, thus far only RE nuclear spin systems have been used for storing optical pulses using spin states~\cite{Ferguson2016,Laplane2017,Kutluer2017,Seri2017}, based on the nuclear quadrupole states of either Pr$^{3+}$ or Eu$^{3+}$ ions. While these non-Kramers RE ions with quenched electronic spin provide excellent memory times, the purely nuclear states limit the memory bandwidth to $<10~\textrm{MHz}$. RE ions with non-zero electronic spins, so-called Kramers ions, could potentially provide a solution to the bandwidth limit, provided that long coherence times can be engineered in such electronic spin systems.

Recently, it has been shown that such long spin coherence times can be found in some Kramers RE ions with electron spin ~$\textbf{S}$~\cite{Kindem2018,Rancic2018,Rakonjac2018,Ortu2018} and non zero nuclear spin  $\textbf{I}$ by exploiting the hyperfine coupling $\textbf{S} \cdot \textbf{A} \cdot \textbf{I}$, where $\textbf{A}$ is the hyperfine tensor.
In Refs~\cite{Rakonjac2018,Ortu2018} it was particularly shown that at zero applied magnetic field, an anisotropic hyperfine interaction leads to strong mixing of $\textbf{S}$ and $\textbf{I}$, resulting in highly hybridized electron-nuclear states with zero first-order Zeeman (ZEFOZ) effect~\cite{Fraval2004}. Using this feature we showed simultaneous long optical and spin coherence times of $T^{o}_2 = 180$~\textmu s and $T^{s}_2 = 1.5$~ms, respectively, in \ybiso~\cite{Ortu2018}. We also showed that, while the hybridized states are insensitive at first order to slowly fluctuating magnetic DC fields (at the ZEFOZ point), the magnetic AC transition moment between the hybridized states remains electronic, resulting in high Rabi frequency and fast operations. These results are promising for broadband and long-duration optical quantum memories, but thus far there has been no demonstration of optical storage using spin states in any Kramers RE ion system.

In this Letter we demonstrate an optical memory using the hybridized electron-nuclear states at zero field in \ybiso, based on the atomic frequency comb (AFC) memory scheme~\cite{Afzelius2009a}. We reach a spin storage time of 1.2~ms and a bandwidth of 10 MHz, which is only limited by the current optical Rabi frequency. In addition, we show efficient optical AFC echoes with delays one order of magnitude longer than previously achieved in any Kramers system, which we attribute to reduced superhyperfine coupling in the zero-field ZEFOZ point.

The AFC memory is based on a $\Delta$-periodic structure of highly absorbing peaks within an inhomogeneously broadened optical transition \cite{Afzelius2009a}. An input pulse then produces an optical AFC echo, with a delay of $1/\Delta$. The AFC echo process allows high temporal multimode storage, provided that $1/\Delta$ is much longer than the input pulse duration~\cite{Jobez2016}. To achieve on-demand read out, an optical control pulse can be applied before the AFC echo, thereby converting the optical coherence into a spin coherence. This spin-wave memory~\cite{Afzelius2010,Gundogan2015,Jobez2015,Seri2017} is read out by applying another control pulse after a time $T_\text{S}$, which results in an output pulse with a storage time of $T_\text{M} = 1/\Delta + T_\text{S}$. The spin-wave storage time can be extended to the spin coherence time $T_2^\text{s}$ by applying a spin echo sequence~\cite{Jobez2015}.

The memory puts specific demands on the atomic system. It requires an excited state coupled to two spin states, a so-called $\Lambda$ system, where the optical memory bandwidth is ultimately limited by the spin-state energy split~\cite{Afzelius2009a}. Efficient AFC echoes and long $1/\Delta$ delays require an optically deep and high resolution comb, which in turn requires a long optical coherence time and efficient optical pumping. Finally, long memory lifetime requires a long $T_2^\text{s}$ and efficient spin manipulation through microwave (MW) pulses.

The optical transition we use in \ybiso{} connects the lowest crystal field levels in the electronic \gstate{} ground and \estate{} excited states, at 978.854~nm (in vacuum) for site~II in \yso{} \cite{Welinski2016}. The highly anisotropic hyperfine tensors splits both levels into four non-degenerate hyperfine states at zero applied magnetic field \cite{Tiranov2018a}, as shown in~\figref{fig:exp_setup}a. Many different combinations of transitions could be used as the $\Lambda$ system for optical storage, with corresponding microwave frequencies from 529 to 3026 MHz. Here we focus on the particular $\Lambda$ system formed by the $\nu_1$ and $\nu_2$ transitions, see~\figref{fig:exp_setup}a, with a spin transition at $\nu_{\text{MW}}=655$~MHz, which has the required optical and spin coherence times \cite{Ortu2018}. A crucial first step towards optical spin-wave storage is then to demonstrate efficient optical pumping and optical AFC echoes.

Optical pumping of all hyperfine states requires addressing transitions involving all four hyperfine states. Our setup (\figref{fig:exp_setup}d) is based on two lasers (master and slave), where the master is locked to a high-finesse cavity at the frequency of the $\nu_1 = 306263.0$~GHz transition. The slave is locked to the master laser with an offset $\nu_2 = \nu_1 + 0.6547$~GHz using an optical phase lock loop (PLL). Additionally the slave laser addresses the $\nu_3 = \nu_2 + 6.2594$~GHz and $\nu_4 = \nu_2 + 7.0762$~GHz transitions by phase modulation (\figref{fig:exp_setup}a and d).

\begin{figure}[ht]
	\includegraphics[width=\linewidth]{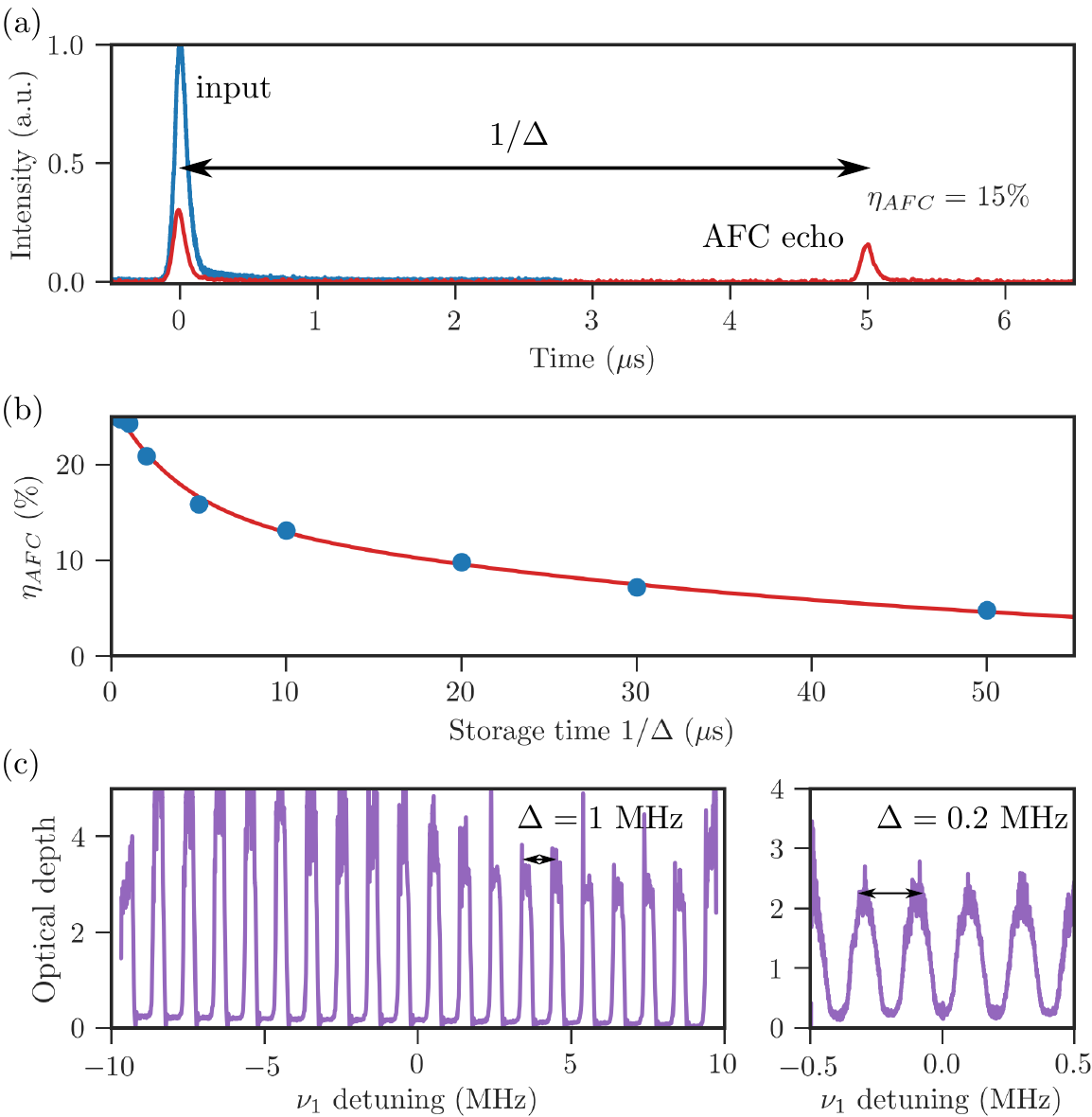}
	\caption{(color online) \textbf{Optical AFC echoes}. 
	\textbf{(a)} Example of an optical AFC echo for a delay of $1/\Delta=5$~\textmu s, with an efficiency of $\eta_{AFC} = 15\%$. The input pulse duration was 100~ns. 
	\textbf{(b)} AFC echo efficiencies as a function of delay $1/\Delta$ (see text for fit model). 
	\textbf{(c)} Two examples of measured AFCs for $1/\Delta=1$~\textmu s and $1/\Delta=5$~\textmu s, in terms of optical depth $d$ as a function of frequency detuning on the transition $\nu_1$.}
	\label{fig:exp_results}
\end{figure}

The \yso{} crystal is doped with 5~ppm of \ybi{}, with an isotopic purity of $\approx 95$\%. The optical inhomogeneous broadening is 1.3~GHz, hence there is clear overlap between optical transitions (\figref{fig:exp_setup}b). This issue can be solved by standard pumping sequences for RE crystals, refereed to as class cleaning \cite{Rippe2005,Lauritzen2012}. By class cleaning for 400 ms with all $\nu_1$-$\nu_4$ frequencies and then state initializing into state $\ket{4}_g$ using frequencies $\nu_2$-$\nu_4$, we obtain the absorption spectrum in \figref{fig:exp_setup}b. The strong optical absorption on the $\nu_1$ transition and the deep holes on the $\nu_2$-$\nu_4$ transitions are evidence of efficient state initialization.

A measurement of the lifetime of the absorption structure showed that it rethermalizes with two different exponential time constants (\figref{fig:exp_setup}c); a faster 36(6)~ms decay and a longer 390(55)~ms decay. The double decay is due the different relaxation rates between hyperfine states, as also observed in $^{145}$Nd$^{3+}$ \cite{ZambriniCruzeiro2018}. The population relaxation lifetimes are significantly longer than in the 10~ppm doped \ybiso{} sample used in Ref.~\cite{Ortu2018}, which indicates that the population relaxation is due to flip-flops between ions in different hyperfine states \cite{ZambriniCruzeiro2017a,ZambriniCruzeiro2018,Car2018}. A more detailed investigation of optical pumping and hyperfine flip-flop relaxation processes in this material will be presented elsewhere~\cite{Welinski2019}.

In a first set of storage experiments the optical AFC echo was studied (\figref{fig:exp_results}), without applying the optical control pulses. The AFC was created on the $\nu_1$ transition, which has the strongest absorption in the $\Lambda$ system. In particular we measured the AFC echo efficiency $\eta_{AFC}$ as a function of the AFC delay $1/\Delta$, which depends strongly on the contrast and shape of the comb. The ideal shape of the comb teeth is squarish, with an optimal comb finesse given only by the maximum optical depth~\cite{Bonarota2010}. To create such combs we use the optical pumping method presented in Ref.~\cite{Jobez2016}, which was specifically designed to create optimal AFCs over a large frequency bandwidth. The AFC bandwidth was set to 20~MHz, and the 60~ms long comb preparation sequence directly followed the state initialization and class cleaning sequences.

An example of an optical AFC echo is shown in \figref{fig:exp_results}a, and the echo efficiencies are plotted in \figref{fig:exp_results}b. At the shortest delay of $1/\Delta =1$~\textmu s, the AFC echo reaches a combined storage and retrieval efficiency of $\eta_{AFC}=24$\%. The associated comb is shown in \figref{fig:exp_results}c, and it features a high contrast, squarish shape that is close to optimal. The theoretical efficiency for an optimal comb for an optical depth of $d=4$ is 32\% \cite{Bonarota2010}. But the experimental comb at $1/\Delta =1$~\textmu s has a background optical depth of about $d_0 \approx 0.3$, which reduces the efficiency to $0.32\exp(-d_0) = 24$\% \cite{Riedmatten2008}, consistent with the experimental data. Higher efficiencies should be achieved by improving the optical pumping, which requires a lower \ybi{} concentration to reduce the flip-flop rate \cite{ZambriniCruzeiro2017a}, and  using optical cavities \cite{Sabooni2013,Jobez2014}.

The decay of the AFC echo as a function of $1/\Delta$ is due to a reduction in contrast and a small deviation in shape from the optimal square one, as exemplified for $1/\Delta=5$~\textmu s in \figref{fig:exp_results}c. The decay curve can be fitted using the formula $\exp(-4/(\Delta \cdot T_2^{'}))$, where ideally the $T_2^{'}$ is the optical coherence time \cite{Jobez2016}. However, shorter $T_2^{'}$ are typically obtained due to technical noise such as laser coherence time limitations and cryostat vibrations. The data in \figref{fig:exp_results}b shows a double exponential decay with $T_2^{'} = 15$ and 165~\textmu s, respectively, while the optical coherence time in this material we measured to be as long as 600~\textmu s with photon echoes. We believe the laser spectrum to be the main limitation to the observed decay constants.

The timescale of the AFC delays shown in \figref{fig:exp_results}b is up to two orders of magnitude longer than previously achieved delays ($0.1-1$~\textmu s) in RE ions with electronic spin degrees of freedom, such as in Nd$^{3+}$ \cite{Usmani2010,Clausen2011} or Er$^{3+}$ \cite{Craiciu2019} doped crystals. Those short decays have been explained \cite{Usmani2010,Craiciu2019} by invoking superhyperfine interaction between the RE electronic spin and the nuclear spin of Y$^{3+}$ ions in the host, which causes spectral nuclear spin-flip sidebands \cite{Wannemacher1991} and effectively enlarges the homogeneous linewidth of the RE ion. The efficient AFC echoes at long delays suggest that superhyperfine interaction is strongly suppressed at the zero-field ZEFOZ point. The delays are similar to those achieved in the purely nuclear RE spin systems Eu$^{3+}$ \cite{Jobez2016} and Pr$^{3+}$ \cite{Seri2018}, which yet again highlights the interest of the hybridization of the electronic and nuclear spins at zero field in \ybiso{}.

\begin{figure}[ht]
	\includegraphics[width=\linewidth]{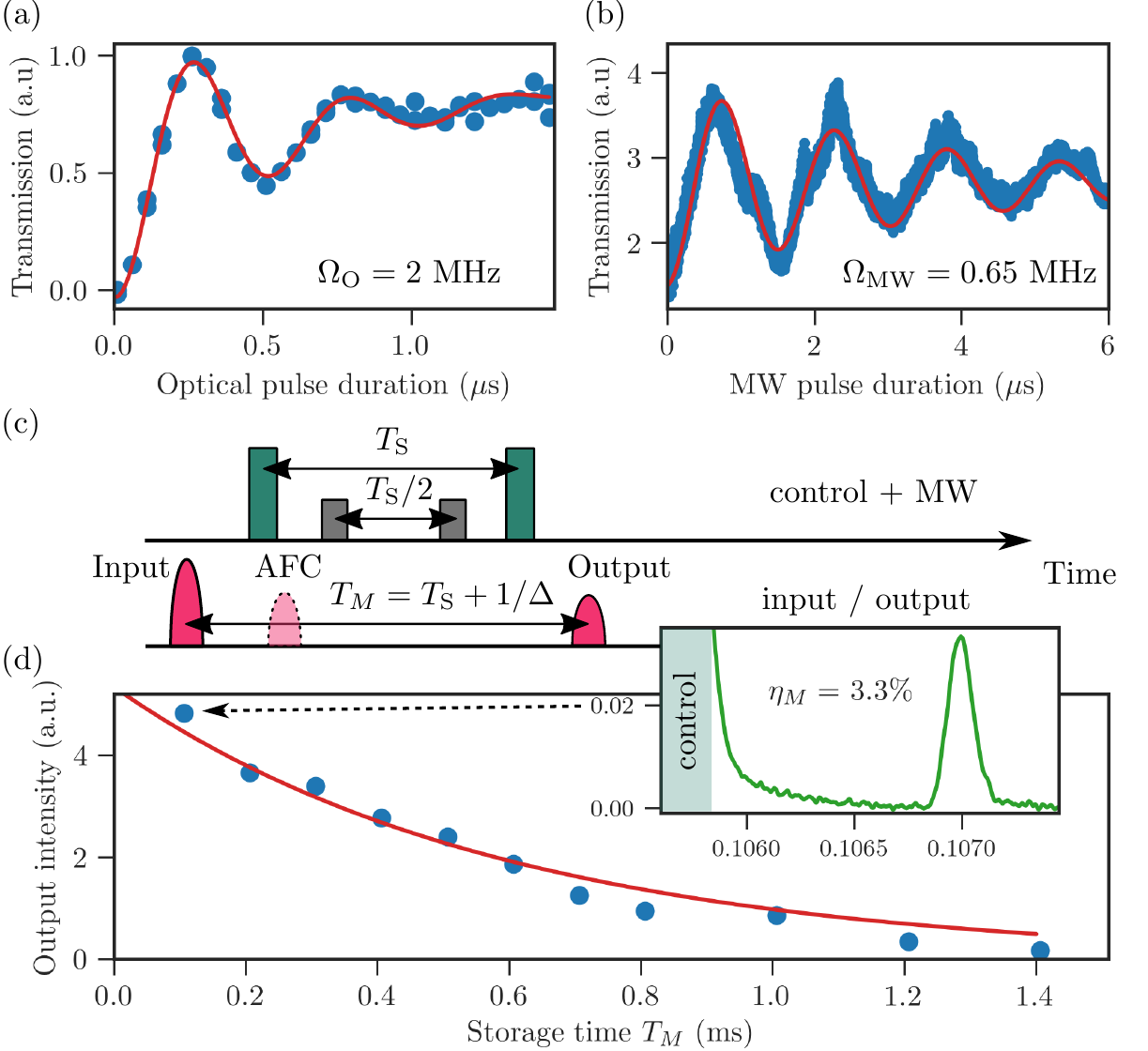}
	\caption{(color online) \textbf{AFC spin-wave storage results}. 
		\textbf{(a,b)} Measured Rabi oscillations on the optical and MW transitions (see \SM~\cite{Businger2019_SM}). 
		\textbf{(c)} Pulse and timing sequence of the AFC spin-wave storage experiment, including input/output pulses (transition $\nu_1$), control pulses (transition $\nu_2$) and MW pulses (transition $\nu_{\text{MW}}$). Note that for a perfect control pulse the AFC echo at $1/\Delta$ is completely suppressed. 
		\textbf{(d)} The intensity of optical output pulse as a function of the total memory storage time $T_\text{M}$, with $1/\Delta=7$~\textmu s. The data was fitted to the function $\exp(-2T_S/T_2^s)$, resulting in a spin coherence time of $T_2^s=1.2(2)$~ms. Inset: Example of output pulse trace for $T_\text{M} = 107$~\textmu s.}
	\label{fig:SW}
\end{figure}

We now turn to the optical spin-wave storage experiments, which in addition require coherent manipulation of the optical and microwave transitions, see \figref{fig:SW}. Optically one requires an efficient population transfer of the control pulse on the $\nu_2$ transition, see \figref{fig:exp_setup}a, which has a dipole moment about 3 times weaker than the $\nu_1$ transition (see \SM~\cite{Businger2019_SM}). On the $\nu_1$ transition we achieved an optical Rabi frequency of $\Omega_{\text{O}} = 2.0$~MHz, see \figref{fig:SW}a, which gives an estimated Rabi frequency of $\Omega_{\text{O}} = 0.6$~MHz on $\nu_2$. The resulting $\pi$-pulse duration of 0.8 $\mu$s implies an efficient population transfer over less than its Fourier limited bandwidth of $1.2$~MHz, clearly insufficient for the 10~MHz bandwidth of the 100~ns input pulse. To increase the bandwidth over which efficient population transfer can be achieved, one can employ longer frequency-chirped adiabatic pulses \cite{Minar2010}. In this case we employ hyperbolic-square-hyperbolic (HSH) pulses \cite{Tian2011} of duration 5~\textmu s, chirped over 10 MHz, which resulted in a measured population transfer efficiency of about 90\% per HSH pulse. To reach larger memory bandwidths would require higher control pulse Rabi frequencies (see \SM~\cite{Businger2019_SM}).

On the microwave transition $\nu_{\text{MW}}$, the stored spin coherence induced by the optical control pulse will dephase with the inverse of the inhomogeneous spin linewidth $1/\Gamma_{\text{MW}}$. We measured a spin linewidth of $\Gamma_{\text{MW}}=0.7$~MHz (see \SM~\cite{Businger2019_SM}), which practically makes it impossible to read out the memory given the duration of the optical control pulse. But the spin coherence can be rephased with a spin echo sequence \cite{Heinze2013,Jobez2015}, in this case a pair of MW $\pi$ pulses greatly extends the storage time. Using a simple coil wrapped around the crystal we reached a Rabi frequency of $\Omega_{\text{MW}} = 0.65$~MHz, see \figref{fig:SW}b. We note that with respect to pure nuclear non-Kramers systems, the spin linewidth is significantly larger. But this is more than compensated for by the electronic spin transition moment, as reflected by the large Rabi frequency. Using adiabatic MW pulses with a duration of 10~\textmu s and a chirp bandwidth of 3~MHz, we readily reach an estimated transfer efficiency of $>$95\%.

The final AFC spin-wave storage experiment data is shown in \figref{fig:SW}d, as a function of the total storage time $T_\text{M} = 1/\Delta + T_\text{S}$. Memory output pulses were detectable beyond 1~ms and the memory lifetime is consistent with previous measurements of the spin coherence time in \ybiso{} \cite{Ortu2018}. It should be emphasized that we reach spin-wave storage times presently only achieved in Eu$^{3+}$:Y$_2$SiO$_5$ crystals using nuclear states. For the shortest spin storage time of $T_\text{S} = 100$~\textmu s, the total memory efficiency was $\eta_\text{M}=3.3\%$. While the AFC echo efficiency could be understood by the optical depth, see \figref{fig:exp_results}, the total memory efficiency falls short of our predictions by a factor of 4, given the optical and MW control pulse efficiencies given above (see \SM~\cite{Businger2019_SM}). Possibly these were overestimated and/or their phase coherence was not sufficient.

To summarize, in this Letter we have demonstrated storage of optical pulses using the electronic-nuclear hyperfine states in \ybiso{}, which constitutes the first demonstration of spin-wave storage in any RE ion with electronic spin. Moreover, the AFC echo delay (which is related to temporal multimode capacity) and the total spin-wave storage time reach similar performance as in the pure nuclear \euiso:Y$_2$SiO$_5$ system, but with 5 times larger optical bandwidths \cite{Laplane2016a}. 

To conclude, we briefly discuss current limitations and future prospects of the memory. The memory bandwidth of 10 MHz is currently limited by the 0.6~MHz Rabi frequency of the control pulses. Simulations show (see \SM~\cite{Businger2019_SM}) that a 2~MHz Rabi frequency could increase the memory bandwidth to 100~MHz. Other $\Lambda$ systems could reach such Rabi frequencies using different polarization modes. To efficiently excite microwave transitions at $>1$~GHz in these $\Lambda$ systems, one can use lumped-element MW cavities \cite{Chen2016,Angerer2016}. The optical Rabi frequency can also be greatly increased by using laser-written waveguides \cite{Corrielli2016,Seri2018} and could potentially allow memory bandwidths in the range of 100~MHz to 1~GHz. This would facilitate interfacing with quantum photon pair sources \cite{Saglamyurek2011,Clausen2014a}, and possibly allow interfacing with quantum dot single photon sources~\cite{Grange2017,Kirsanske2018,Schweickert2018}. There is also the prospect of greatly increasing the AFC multimode capacity, which would increase the rate of a quantum repeater~\cite{Simon2007}. By increasing the bandwidth and by achieving AFC echo delays only limited by the long optical coherence time in \ybiso{}, the temporal multimode memory capacity could potentially reach 1000 modes. An important future step will be to demonstrate storage of weak coherent states, a strong requirement for a quantum memory. This will crucially depend on the efficiency with which one can optically pump ions out of the storage state, to decrease unwanted photon emission noise due to the control pulses. Recent work has shown that optical pumping of hyperfine states can be particularly efficient \cite{ZambriniCruzeiro2018}.

\section*{ACKNOWLEDGEMENTS}

We acknowledge funding from FNS Research Project No 172590, EUs H2020 programme under the Marie Sk\l{}odowska-Curie project QCALL (GA 675662), IMTO Cancer AVIESAN (Cancer Plan, C16027HS, MALT).

\clearpage
\newpage


\section*{Supplementary material}

# Supplemental Materials: "Optical spin-wave storage in a solid-state hybridized electron-nuclear spin ensemble"

M. Businger,[1, *] A. Tiranov,[1, †] K. T. Kaczmarek,[1] S. Welinski,[2, ‡]
Z. Zhang,[2] A. Ferrier,[2, 3] P. Goldner,[2] and M. Afzelius[1, §]

[1]*Department of Applied Physics, University of Geneva, CH-1211 Genève, Switzerland*
[2]*Chimie ParisTech, PSL University, CNRS, Institut de Recherche de Chimie Paris, 75005 Paris, France*
[3]*Faculté des Sciences et Ingénierie, Sorbonne Université, UFR 933, 75005 Paris, France*

### Crystal

Our sample is a $^{171}$Yb$^{3+}$:Y$_2$SiO$_5$ crystal with 5 ppm doping concentration and with an Yb$^{3+}$ isotope purity of 95%. The inhomogeneous broadening on the optical $^2$F$_{7/2}$(0) $\longleftrightarrow$ $^2$F$_{5/2}$(0) transition was measured to be 1.3 GHz which is noticeably larger than 0.56 GHz measured in the 10 ppm doped sample [1]. We attribute this to the imperfections during the growing process of the crystal. The crystal was grown using the Czochralski method and cut along orthogonal polarization extinction axes of the Y$_2$SiO$_5$ crystal usually denoted as $D_1$, $D_2$ and $b$, where $b$ is the crystallographic $C_{2h}$ crystal symmetry axis. The dimensions along $D_1 \times D_2 \times b$ are $3.5 \times 4.5 \times 12$ mm, respectively. The crystal was polished for light propagating along the $b$ axis. The polarization of the light was along the $D_2$ axis to maximize the absorption for the $\nu_1$ transition.

### Population relaxation

The population relaxation was probed using the optical spectral hole burning technique [2]. As mentioned in the main text, several decay constants were observed which can be interpreted as relaxation processes involving various hyperfine transitions [3]. The longest relaxation time was measured to be about 400 ms, while the shortest is 40 ms long. From the decay of the sidehole and the antihole structure we can attribute the fast decay to the relaxation of the $|4\rangle_g \leftrightarrow |3\rangle_g$ and the $|1\rangle_g \leftrightarrow |2\rangle_g$ transitions coming from the spin cross-relaxation. The long decay can be attributed to the relaxation of the $|4\rangle_g \leftrightarrow |2\rangle_g$ and the $|1\rangle_g \leftrightarrow |3\rangle_g$ transitions.

### Spin Rabi frequency

To address the spin $|4\rangle_g \leftrightarrow |3\rangle_g$ transition we use an unmatched inductive copper coil wrapped around the crystal with 6 turns aligned with the $b$ axis of the crystal. To drive the transition the Agilent ESG signal generator is used together with a 4 W microwave amplifier. Using the impedance-unmatched coil with an inductance of $\approx$ 200 nH an induced oscillating magnetic field at

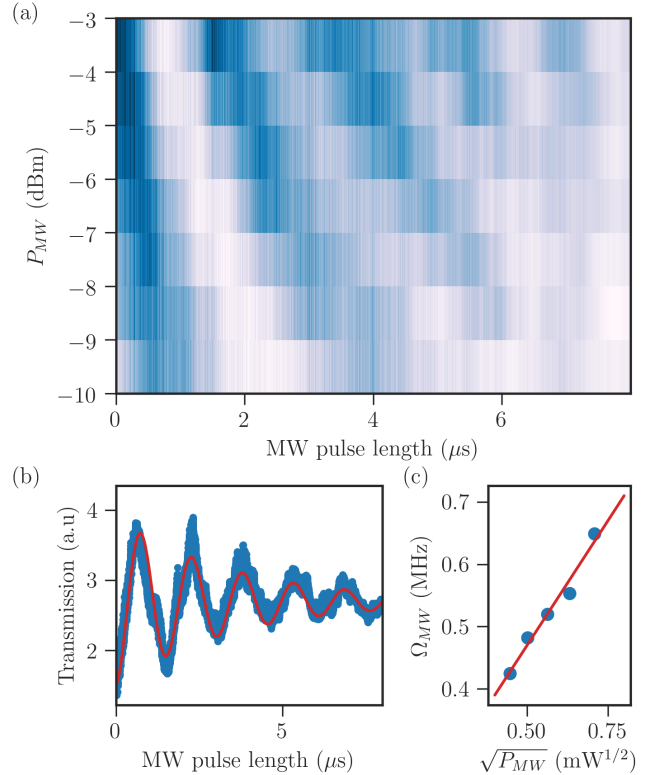

FIG. 1. (color online) **Coherent driving of the spin transition**. **(a)** The oscillation of the transmitted optical signal synchronised with the 10 µs microwave pulse. The measurement is done for various microwave power levels at the input of the microwave 33 dB amplifier. **(b)** Temporal trace measured for -3 dBm input power. The measured oscillation corresponds to a Rabi frequency of $\Omega_{\mathrm{mw}} = 0.65$ MHz. **(c)** Measured Rabi frequency for various input power levels of the microwave pulse. The square root $\sqrt{P_{mw}}$ dependence of the measured frequency confirms the coherent nature of the observed spin manipulation.

655 MHz on the order of $10 - 20$ µT was achieved.

The coherent operation on the spin $|4\rangle_g \leftrightarrow |3\rangle_g$ transition was tested using the optical probe pulse synchronised with a microwave square pulse (FIG. 1). By looking at the absorption of the probe signal a Rabi oscillation with up to $\Omega_{\mathrm{MW}} = 0.65$ MHz frequency was measured. This is in accordance with the estimated microwave power cou-

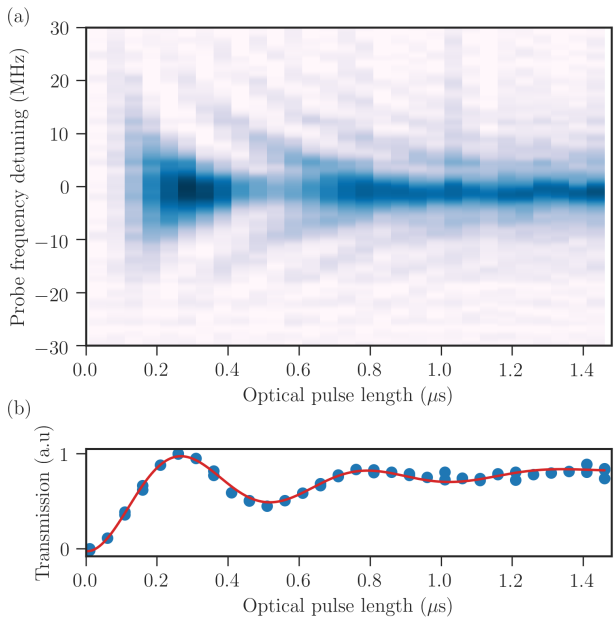

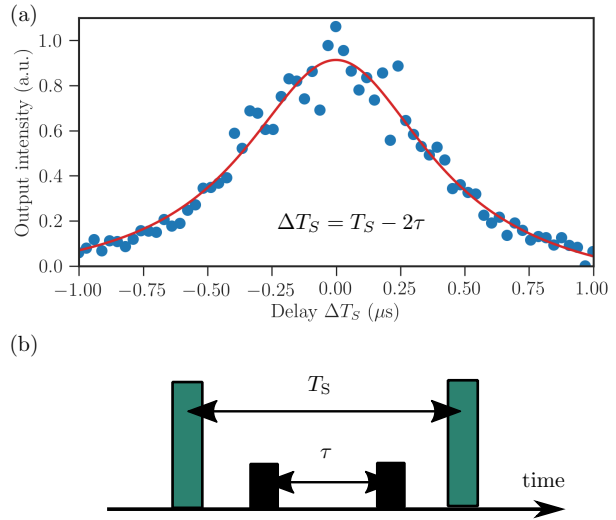

FIG. 2. (color online) **Coherent driving of the optical transition**. (a) Optical transmission for $|4\rangle_g \leftrightarrow |1\rangle_e$ measured right after the strong optical control pulse. The transmission is measured by scanning the frequency of the probe pulse for different durations of the control pulse. Clear oscillations are visible due to the coherent driving of the optical transition. The faster oscillations are visible for detuned spectral components due to the square temporal shape of the optical pulse and a faster Rabi frequency for off-resonant excitations. (b) Lower figure shows the transmission of the probe pulse for the central part around zero detuning as a function of the pulse length. The oscillation corresponds to a Rabi frequency of $\Omega_O = 2$ MHz and was measured for the maximum optical power of $\approx 300$ mW before the cryostat.

FIG. 3. (color online) **Spin inhomogeneous linewidth measurement**. (a) The intensity of spin-wave storage signal measured for different temporal positions of the second optical transfer $T_S$ for fixed temporal separation between two microwave pulses $\tau$. The delay from the optimal value of $\Delta T_S = T_S - 2\tau$ leads to a fast dephasing in the spin transition due to the finite inhomogeneous broadening. By fitting to the Eq. (1) we extract the inhomogeneous linewidth of $\Gamma_{MW}$ = 0.73(4) MHz. This value fits well with measurements done with Raman heterodyne signal. (b) The pulse sequence used for this experiment with timing indications.

pled to the unmatched coil, taking into account the magnetic dipole moment of the transition $g\mu_B$ with $g = 6.06$ and $\mu_B/h = 13.99$ GHz/T for a microwave excitation along the $b$ axis.

To generate efficient $\pi$-rotations in the microwave domain we used sech/tanh adiabatic-shape chirped pulses. This allows one to achieve efficient population inversion over the entire spin inhomogeneous linewidth of the spin ensemble. For this we use the IQ-modulation channels of the Agilent ESG signal generator which is controlled by an arbitrary function generator (Tektronix 3022C).

The efficiency of the $\pi$-pulse was estimated by probing the population using the optical probe pulse and measuring the population change before and after several pairs of $\pi$ pulses. An efficiency of up to 97(1)% was deduced from this measurement. We attribute this imperfection to the spatial inhomogeneity of the Rabi frequency induced by the copper coil along the crystal.

### Spin rephasing

The inhomogeneous broadening of the hyperfine transition was previously estimated using Raman heterodyne scattering to be about $\Gamma_{MW} = 730$ kHz. This value is an order of magnitude bigger than for the hyperfine transitions of non-Kramers ions. This leads to a much faster dephasing of the induced spin coherence, which happens in $\approx 500$ ns. Such a fast dephasing imposes a strong constraint on the separation between the two control field pulses to observe the three-level AFC echo after transferring from the spin state.

This issue can be solved by applying a pair of $\pi$ pulses on the microwave transition in order to rephase the induced spin coherence (see FIG. 5c). The distance between the pulses $\tau$ has to match the timing between the control field pulses $T_S$ such that $\tau = T_S/2$. To probe the dephasing process one can measure the efficiency of the rephased signal as a function of the delay between the $\pi$ pulses $\eta_M(\tau)$. If we assume a Gaussian profile for the spin inhomogeneous broadening, the decay of the stored echo is given by

$$\eta_M(\tau) = A \exp\left\{-\frac{\pi^2 \Gamma_{MW}^2 (T_S - 2\tau)^2}{2\ln 2}\right\}, \quad (1)$$

where $A$ is the constant including the efficiency of the

TABLE I. Measured relative optical oscillator strengths for $^{171}\text{Yb}^{3+}$:$Y_2SiO_5$ optical $^2F_{7/2}(0) \longleftrightarrow ^2F_{5/2}(0)$ transition for crystallographic site II measured for light polarized along the $D_2$ axis using the spectral holeburning technique. The lambda system used for the spin-wave storage experiment is indicated by bold.

|         | $|1\rangle_e$ | $|2\rangle_e$ | $|3\rangle_e$ | $|4\rangle_e$ |
|---------|------|------|------|------|
| $\langle 1|_g$ | 0.15 | 0.06 | 0.08 | 0.71 |
| $\langle 2|_g$ | 0.06 | 0.19 | 0.71 | 0.04 |
| $\langle 3|_g$ | **0.07** | 0.71 | 0.16 | 0.06 |
| $\langle 4|_g$ | **0.72** | 0.04 | 0.05 | 0.19 |

two-level AFC $\eta_{\text{AFC}}$ and the finite efficiency of the transfer $\eta_t$ and microwave $\eta_{\text{MW}}$ pulses.

The results are depicted in FIG. 3 together with the fit by Eq. (1). The extracted FWHM of the inhomogeneous broadening of the spin transition of $\Gamma_{\text{MW}} = 0.73(4)$ MHz is in accordance with the previous estimation.

### Optical branching ratio

The optical branching ratio TABLE I was measured for a sample with a 10 ppm concentration. The light was polarized along the $D_2$ axis. The spectral hole burning technique was used to extract hole/antihole amplitudes for every optical transition. For this the spectral hole was prepared at $\nu_1$ laser frequency, while the probe frequency was scanned over a wide range (10 GHz) to measure the whole optical absorption spectra. The hole/antihole structure was further extracted for the measured spectra.

The optical inhomogeneous broadening for that sample is 0.56 GHz, which results in a low overlap between different spectral classes. In this case the amplitudes for most of the holes/antiholes represent the absorption coefficient for certain optical transitions. This way, the optical branching ratios for the given ground states connected to different excited states were extracted. Further normalization was used to reconstruct the whole $4 \times 4$ table (TABLE I).

### Optical control pulse bandwidth

Using $\approx 300$ mW power for the optical control field we measured the Rabi oscillation of $\approx 2$ MHz for the $|4\rangle_g \leftrightarrow |1\rangle_e$ optical transition (FIG. 2). This value is limited by the low oscillator strength of the transition and, in principle, can be enhanced by an external optical cavity or a strong light confinement using waveguide structures. An order of magnitude increase of the Rabi frequency has been demonstrated previously using those methods [4, 5].

By using the optical branching ratio table (TABLE I) we estimate the Rabi frequency of $|3\rangle_g \leftrightarrow |1\rangle_e$ to be $\approx$ 0.6 MHz. Hence, this is the maximum Rabi frequency we currently can achieve with the optical control pulse.

To increase the bandwidth over which one can achieve efficient population (or coherence) transfer one can employ frequency-chirped adiabatic pulses, as discussed in detail in Ref. [6]. In this experiment we use hyperbolic-square-hyperbolic (HSH) pulses [7], which were specifically designed to achieve efficient transfer over a large bandwidth with respect to the available Rabi frequency. Using a simple Bloch calculation we simulated the population transfer as a function of detuning, for a given Rabi frequency. The input parameters were the HSH pulse duration $T$, Rabi frequency $\Omega_O$, and the total frequency chirp bandwidth $\Gamma$ of the HSH pulse. Note that $T$ refers to the central part of the HSH pulse, which is the period during which the Rabi frequency is constant. The smoothly rising and falling edges of the HSH pulse are much shorter for the pulses used during this work, and they have negligible effect on the resulting transfer efficiency.

The average efficiency $\eta_t$ of the HSH pulse is then calculated by averaging the population transfer over the chirp bandwidth $\Gamma$. In the adiabatic limit, the HSH pulse performs equally well over this entire bandwidth, hence it has a top-hat profile as a function of detuning [7]. However, if the pulse is not adiabatic the transfer function will have a smoother Gaussian shape, and the average efficiency will be lower than the peak efficiency at zero detuning, see for instance [8].

In FIG. 4, average transfer efficiency $\eta_t$ is plotted as a function of $\Gamma$, for Rabi frequencies ranging from $\Omega_O/(2\pi)=0.6$ MHz to $\Omega_O/(2\pi)=2$ MHz. The pulse duration was set to $T = 5 \mu$s. For lower chirp bandwidths, the behavior is complex with the appearance of a maximum efficiency. We attribute this to the effect of non-adiabaticity discussed above. In the regime of large chirp bandwidth, however, the efficiency drops as expected when the chirp bandwidth is increased. It is also clear that for our Rabi frequency of $\Omega_O/(2\pi)=0.6$ MHz, the maximum efficiency is reached for about $\Gamma/(2\pi)=10$ MHz, and increasing the bandwidth to $\Gamma/(2\pi)=20$ MHz significantly reduces the efficiency. Note that the total AFC spin-wave memory efficiency scales as $\eta_t^2$.

The HSH pulse is an extension of the well-known adiabatic population inversion pulse called hyperbolic secant (SECH) pulse, for which one can analytically solve the Bloch equations [8]. The HSH pulse is a time stretched SECH pulse that has larger effective pulse area for the same pulse duration. As a consequence it has a higher population transfer efficiency for finite pulse duration and Rabi frequency, with respect to the basic SECH pulse. But as it is derived from a SECH pulse, the analytical results of SECH pulses still apply. In Ref. [6] the average population transfer efficiency was calculated for a SECH

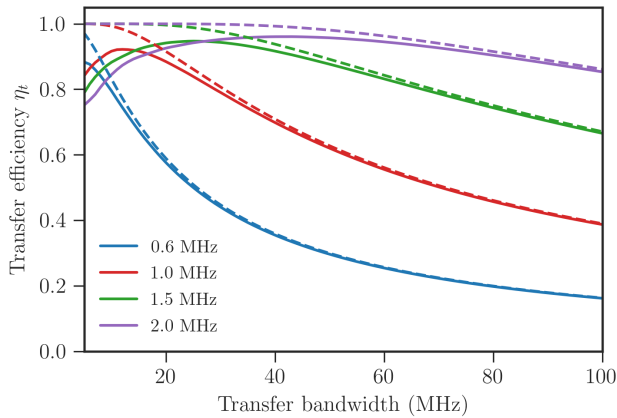

FIG. 4. (color online) **Theoretical optical control pulse efficiency**. Solid lines show the simulated average transfer efficiency $\eta_t$ as a function of the HSH bandwidth $\Gamma/(2\pi)$, for Rabi frequencies $\Omega_O/(2\pi) = 0.6, 1.0, 1.5, 2.0$ MHz. Dashed lines show the corresponding efficiency calculated using Eq. (2).

pulse, for a given set of pulse parameters. Our simulations show that the HSH pulse has the same scaling as a function of these parameters, with the theoretical efficiency

$$\eta_t^{\text{HSH}} \approx 1 - \exp\left(-0.5 \frac{\pi T \Omega_O^2}{\Gamma}\right). \quad (2)$$

The formula was simplified under the approximation that $\Gamma >> \Omega_O$. The formula has the same functional form as for SECH pulses, but we had to fit the pre-factor of 0.5 appearing in the exponential to obtain exact agreement with the Bloch simulations. The theoretical efficiencies calculated using Eq. (2) are shown as dashed lines in FIG. 4. The quadratic dependence on the Rabi frequency $\Omega_O$ explains the strong influence of the Rabi frequency on the achievable bandwidth.

### Two-level AFC

By varying the frequency separation of the comb teeth $\Delta$ one can probe the quality of the prepared comb structure. For this the two-level AFC echo experiment was performed for various storage times (see FIG. 2b in the main text). Due to the finite width of the created structure the intensity of the echo reemission will degrade for longer $1/\Delta$ storage times with characteristic decay time $T_2^*$ characterizing the spectral comb

$$\eta_{\text{AFC}} = \eta_{\text{AFC}}^{\text{opt}} \exp\left\{-\frac{4}{T_2^* \Delta}\right\}, \quad (3)$$

where $\eta_{\text{AFC}}^{\text{opt}}$ is the maximum efficiency for a given optical depth of the comb $d$, optimal finesse $F$ and background absorption $d_0$ given by

$$\eta_{\text{AFC}}^{\text{opt}} = \left(\frac{d}{F}\right)^2 \text{sinc}^2\left(\frac{\pi}{F}\right) e^{-d/F - d_0}. \quad (4)$$

The optimal finesse is $F = \pi/\arctan(2\pi/d)$ and the ultimate limit for the decay time $T_2^*$ is given by the coherence lifetime of the optical transition $T_2^o$. The highest efficiency for a given optical depth $d$ is obtained for square peaks, where the intrinsic dephasing from the absorption peak shape is given by the second term in the Eq. (4). In this scenario the echo efficiency is bounded by 54% due to reabsorption (the last term in the Eq. (4)). This limit can be overcome by forcing the reemission in the backward direction or by placing the crystal inside an impedance-matched cavity to reach a theoretical limit of 100% [5, 9].

The effective decay $T_2^*$ extracted from the experiment contains two decay constants of $\approx 16\,\mu s$ and $\approx 160\,\mu s$ (see FIG. 2b). These values are still much lower than the measured optical coherence time $T_2^o$ of the optical transition of 600 μs. The effects of the spectral resolution and the laser linewidth are the most probable limitations.

### Experimental setup

Our laser setup is based on two external cavity laser diodes at 978.8 nm (Toptica DLpro and DL100) (see FIG. 5). The first (*master*) laser is stabilized using a custom Fabry-Perot cavity with finesse approaching $\approx 5000$ by the Pound-Drever-Hall technique at $\nu_1 \approx 306263.0$ GHz frequency. The cavity is placed inside a custom vacuum system and is passively stabilized. This laser is used for the preparation of the atomic frequency comb structure at the $|4\rangle_g \leftrightarrow |1\rangle_e$ transition and the creation of the input pulse (see FIG. 5a). This method of locking provides the laser linewidth of $\approx 1$ kHz comparable with the homogeneous linewidth of the optical transition $\Gamma_h = (\pi T_2^o)^{-1}$, which is estimated to be $\approx 2$ kHz.

The second (*slave*) laser is stabilized with a custom optical phase locked loop setup using the external signal generator at $\nu_2 \approx \nu_1 + 0.6547$ GHz optical frequency with respect to the *master* laser [10]. The tapered amplifier (Toptica Boosta) is used to reach 1.2 W optical power level utilized for the control mode. Part of this power is going through the fibre-based Eospace phase modulator to create two additional frequencies $\nu_3 = \nu_2 + 6.2594$ GHz and $\nu_4 = \nu_2 + 7.0762$ GHz to address $|2\rangle_g \leftrightarrow |3\rangle_e$ and $|1\rangle_g \leftrightarrow |4\rangle_e$ optical transitions, respectively. These frequencies are used to polarize the spin ensemble into the $|4\rangle_g$ ground state before the preparation of the AFC structure (see FIG. 5b).

Acousto-optic modulators (AOM) are used to create the necessary light pulse sequence for the storage experiment. It consists of class cleaning (400 ms), population polarization (300 ms), AFC preparation (60 ms)

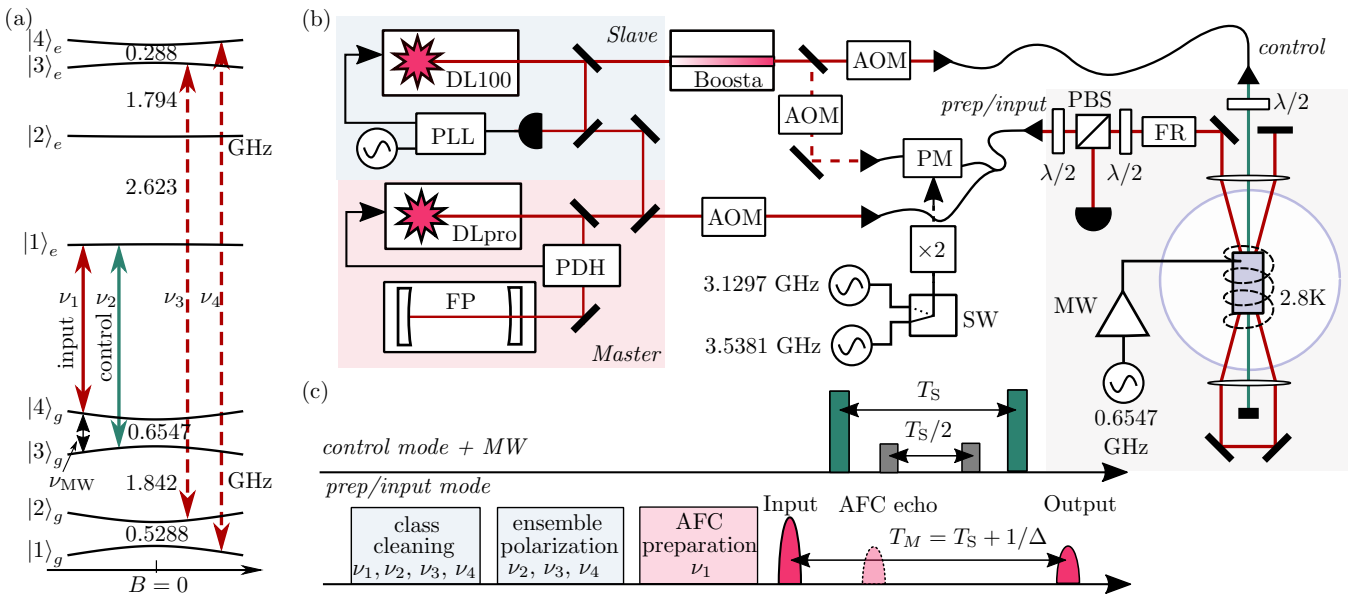

FIG. 5. (color online) **Experimental scheme**. (a) Energy level diagram for optical $^2F_{7/2}(0) \longleftrightarrow {}^2F_{5/2}(0)$ transition for site II of $^{171}$Yb$^{3+}$:Y$_2$SiO$_5$ crystal at zero magnetic field. The $|4\rangle_g \leftrightarrow |1\rangle_e$ ($\nu_1$ frequency) transition if used for a storage while the $|3\rangle_g \leftrightarrow |1\rangle_e$ ($\nu_2$ frequency) transition is used for the optical transfer into the spin state and back. $|1\rangle_g \leftrightarrow |4\rangle_e$ and $|2\rangle_g \leftrightarrow |3\rangle_e$ ($\nu_3$ and $\nu_4$) transitions are used for spin ensemble polarization. (b) Experimental setup (see text for details). FP - Fabry-Perot cavity, PDH - Pound-Drever-Hall module, AOM - acousto-optic modulator, PM - phase modulator, PLL - phase locked loop, $\lambda/2$ - half-wave plate, PBS - polarization beamsplitter, ×2 - microwave frequency doubler, FR - Faraday rotator, SW - microwave switch. (c) The pulse sequence for the optical spin-wave storage experiment. The polarization step takes 300 ms, while the comb is prepared in 60 ms. A waiting time of 4 ms is used between the end of the preparation and the start of the three level AFC sequence. $T_S$ is the time between the control pulses, and $1/\Delta$ is the AFC storage time. The total storage time is $T_M = T_S + 1/\Delta$. The separation between the microwave (MW) pulses is $T_S/2$.

and the storage sequence (see FIG. 5b). The RF signals used to drive the AOMs are generated using an arbitrary waveform generator (Tabor WW1281A), arbitrary function generators (Tektronix 3102C) combined with custom AOM drivers. We use two different optical paths to interact with the sample to spatially separate the input and control modes, which prevents the noise from the strong control mode polluting the echo signal (see FIG. 5c).

After each AOM the light is coupled into a single-mode polarization maintaining fibre and then out-coupled on a separate optical table, where the closed cycle cryogenic cooler (Oxford, PT-405) with an optical access is located (see FIG. 5b). Our $^{171}$Yb$^{3+}$:Y$_2$SiO$_5$ crystal is attached to the cold head using a custom copper support. The support is based on copper braids and suppresses vibrations coming from the cryogenic cold head. This allowed us to run all the experiments without detrimental effects of mechanical vibrations.

The 125 mm lens is used to focus both modes on to the crystal, resulting in a beam width of around 90 µm for the input mode and 150 µm for the control mode. The maximum power in the control mode before the cryostat window is measured to be 300 mW. The maximum power available in the input mode is 5 mW. An additional 125 mm lens is placed after the crystal for spatial mode matching. Two right angle mirrors are used to reflect the input mode symmetrically around the control mode to pass a second time through the crystal, as well an additional mirror is placed to back reflect the input mode. This results in quadruple-passage through the crystal to increase the overall absorption. The optical depth of $\approx 4-5$ at the $|4\rangle_g \leftrightarrow |1\rangle_e$ transition was achieved after the polarization sequence.

The output mode is separated using the free-space Faraday rotator (FR) together with a polarization beam splitter (PBS) and a half-wave $\lambda/2$ plate. The polarization in the quadruple path is set by the half-wave $\lambda/2$ plate parallel to $D_2$ axis of the crystal to maximize the absorption (see FIG. 5b). The separate half-wave $\lambda/2$ plate is used for the control mode to match the polarization of the input mode. The signal is detected using the Thorlabs photodiode with switchable gain corresponding to a bandwidth of 4 MHz.

---

\* These authors contributed equally to this work.
† These authors contributed equally to this work.; Present address: The Niels Bohr Institute, University of Copenhagen, DK-2100 Copenhagen ⊘, Denmark

‡ Present address: Department of Electrical Engineering, Princeton University, Princeton, NJ 08544, USA
§ Email to: mikael.afzelius@unige.ch



\end{document}